\documentclass{article}
\usepackage{graphics}
\usepackage{emulateapj}

\def\etal{{et al.}\thinspace}
\def\beb{\begin{thebibliography}{999}}
\def\eeb{\end{thebibliography}}
\def\bi{\bibitem[]{}}
\def\be{\begin{equation}}
\def\ee{\end{equation}}
\def\bea{\begin{eqnarray}}
\def\eea{\end{eqnarray}}
\def\o{\over}
\newcommand{\gsim}{\raisebox{-0.3ex}{\mbox{$\stackrel{>}{_\sim} \,$}}}
\newcommand{\lsim}{\raisebox{-0.3ex}{\mbox{$\stackrel{<}{_\sim} \,$}}}
\newcommand{\gmf}{\fg}
\newcommand{\fg}{$f_g$ }

\begin{document}

\slugcomment{submitted to ApJL}

\title{Probing the Evolution of Gas Mass Fraction with Sunyaev-Zel'dovich Effect}

\author{Subhabrata Majumdar\altaffilmark{1,2}}
\affil{$^1$Joint Astronomy Programme, Physics Department,
Indian Institute of Science, Bangalore 560012, India}
\affil{$^2$ Indian Institute of Astrophysics, Bangalore 560034, India\\
   email: sum@physics.iisc.ernet.in}

\begin{abstract}
Study of the primary anisotropies of the Cosmic Microwave Background (CMB) can be used to determine the
cosmological parameters to a very high precision.  
The power spectrum of the secondary CMB anisotropies due to the thermal
Sunyaev-Zel'dovich Effect (SZE) by clusters of galaxies, can then be studied, to constrain more cluster specific
properties (like gas mass).
We show the SZE power spectrum from clusters to be a sensitive probe of any possible
evolution (or constancy) of the gas mass fraction. The position of the peak of the SZE power spectrum 
is a strong discriminatory signature of different 
gas mass fraction evolution models. 
For example, for a flat universe, there can be a difference in the $l$ values (of the peak) of {\it as much 
as 3000} between a
constant gas mass fraction  model and an evolutionary one. 
Moreover, observational determination of power spectrum, from blank sky surveys, is
devoid of any selection effects that can possibly affect targeted X-ray or radio
studies of gas mass fractions in galaxy clusters.
\end{abstract}


\section{Introduction}
Clusters of galaxies, being perhaps the largest gravitationally bound structures in the universe, 
are expected to contain a significant amount of baryons of the universe. Moreover, due to their large angular 
sizes, observational estimates of their total mass $M_T$, the gas mass $M_g$ and hence the gas mass fraction
(\fg$={M_g \o M_T}$) are easier.
These estimates can be
used as probes of large scale structure and underlying cosmological models. For example,
the cluster \fg would give a lower limit to the universal baryon fraction $\Omega_b/\Omega_m$. 
Determination of \fg has been done by numerous people (White \& Fabian, 1995; Mohr \etal 1999; Sadat 
\& Blanchard, 2001) and the  values are in agreement within the observational scatter. A point to be noted here is that 
the estimated \gmf  depends on the distance to the cluster (i.e \fg $\propto d_{ang}^{3/2}$ ). Hence, if
\gmf is assumed to be constant, then in principle, one can use the `apparent' evolution of \gmf over a large redshift
range to constrain cosmological models (Sasaki 1996 ).

The question as to whether there is any evolution (or constancy) of gas mass fraction, however, is still 
debatable, with
claims made either way. For example, Schindler (1999) has investigated  a sample of distant clusters
with redshifts between 0.3 to 1 and conclude that there is no evolution of the gas mass fraction. Similar 
conclusion has been drawn by Grego \etal (2000).
On the contrary, Ettori and Fabian (1999) have looked at 36 high-luminosity clusters, and
find evolution in their gas mass fraction (in both sCDM and $\Lambda$CDM universes). See also David \etal,
1995; Tsuru \etal 1997; Allen \& Fabian, 1998; Mohr \etal 1999).
Observations suggest that, though \gmf of massive clusters ($T_e \gsim 5KeV$) appears to be constant, low mass
clusters with shallower potential wells may have lost gas due to preheating and/or post-collapse 
energy input (David \etal 1990, 1995; Ponman \etal 1996; Bialek \etal 2000). It is also well known that ICM 
is not entirely primordial and there is probably continuous infall of gas, thereby, increasing  \fg with time. 
Thus, there is considerable debate regarding the evolution of gas mass fraction.

The intracluster medium (ICM)  has been probed mainly through X-ray observations, but also through the so called 
Sunyaev-Zeldovich effect (Zel'dovich \& Sunyaev, 1969) in the last 
decade (see Birkinshaw 1999 for a review). The SZE effect from
clusters is a spectral distortion of the CMB photons due to inverse Compton scattering by the hot ICM
electrons, with its magnitude proportional to the Compton $y$-parameter, given by 
$y={{k_B \sigma_T}\over{m_e c^2}} \int n_e T_e dl$. Here, $k_B$ is the Boltzman constant, $\sigma_T$ is the
Thomson scattering cross section, $m_e$ is the electron mass,
and  $n_e$ and $T_e$ are the ICM electron density and
the temperature. 
Using bolometric SZE measurements, the \fg  has been obtained for a number of clusters (see Birkinshaw 1999).
Recently, Grego \etal (2000) have made interferometric observations of SZE from a sample of 18
clusters. A major advantage of SZE over X-ray
measurements is, 
SZE does not suffer from the ${(1+z)}^{-4}$  `cosmological dimming', which makes SZE an useful probe of evolution of 
cluster gas mass fraction. 

Other than the targeted SZE observations, non-targeted `blank sky'
surveys of SZE are one of the main aims of future satellite and ground based small angular scale 
observations (Holder \& Carlstrom, 1999; or with AMiBA 
). 
Once the power 
spectrum is extracted from observations, comparison can be made with theory, to constrain
cosmological parameters and relevant cluster scale physics. 
The SZE power spectrum as a cosmological probe has been well studied (Refregier \etal, 1999; Komatsu \& Kitayama,
1999), although its use as a probe of ICM has seldom been looked at.

Keeping such surveys in mind, in this {\it Letter}, 
we look at the SZE power spectrum as a probe of the ICM. We show it to be 
a very sensitive probe of the evolution (or constancy) of \gmf. Measurements of the primary
anisotropy would give us `precise' values of cosmological parameters (like $h,
\Omega_m, \Omega_\Lambda, \Omega_b$). Hence, for our calculations, we assume that we know the 
values of cosmological parameters and do not worry about their effect on the SZE power spectrum. 
Any feature of the SZE power spectrum is attributed to specific cluster physics  (like gas content).
We note that this method of probing \gmf is not biased from any selection effect that can occur while
doing pointed SZE observations of X-ray selected clusters of galaxies, and hence is more desirable.

Current observations of primary CMB anisotropies suggest a flat universe with a cosmological constant 
(Padmanabhan and Sethi, 2000).  For
our calculations, we take a flat universe with $\Omega_m=0.35, \Omega_b=0.05$ and $h=0.65$ as our fiducial model.

The paper is structured as follows. In the next section we discuss the distribution of clusters and model the cluster
parameters. In \S 3, we compute the Poisson and clustering power spectrum from SZE and, finally, we discuss our 
results and conclude in \S 4.. 

\section{The Sunyaev-Zel'dovich effect power spectrum}

\subsection{Distributing the galaxy clusters}
We set up an ensemble of galaxy clusters with masses between 
$10^{13} \le M \le 10^{16}$
M$_{\odot}$, using the abundance of collapsed objects as predicted
by a modified version of the  Press-Schechter (PS) mass function 
(Press \& Schechter 1974) given by Sheth and Tormen (1999). The lower mass cutoff signifies the mass for which one
expects a well developed ICM. 
The gas mass is supposed to sit in the halo potential and are distributed in the same manner.
Note, the results are not sensitive to either lower or upper cutoff.
We probe up to redshifts of 5. Most of the power, however, comes from objects distributed at $z \lsim 1$..

We use the transfer function of Bardeen \etal (1986),
with the shape parameter given by Sugiyama (1995) and the Harrison -
Zel'dovich primordial spectrum to calculate the matter power spectrum
$P_m$(k). The resulting $COBE-FIRAS$ normalised (Bunn \& White, 1997) mass  variance ($\sigma_8$) is 0.9 for our
fiducial model.

\subsection{Modelling the cluster gas}
We assume the ICM to follow a $\beta$-profile with 
$\beta=2/3$ for simplicity. The other physical parameters of the clusters are determined using the virial
theorem and spherical collapse model. We closely follow Colafrancesco \& Vittorio (1994) in our modelling.
We have for the gas density,
$n_e(r) = n_{e,0}{\left(1 + {{r^2}\over{r_c^2}} \right)}^{-3\beta/2}.$
We take the gas to be extended up to $R_v=pr_c$ with $p=10$. 
 The central gas density, $n_{e, 0}$ is given by 
$n_{e,0} = f_g {{2\rho_0}\over{m_p(1+X)}}$
where $X=0.76$ is the average proton mass fraction and $\rho_0$ is the central gas mass density.
To account for the fact that there is a final cutoff in the gas distribution we introduce a Gaussian filter at
the cluster edge $R_v$ given by $n_e(r)\rightarrow n_e(r) e^{-r^2/\xi R_v^2}$, where $\xi = 4/\pi$ is the fudge
factor. 

We parametrize the gas mass fraction as 
\be
f_g = f_{g0} {\left(1+z\right)}^{-s} {\left({{M}\o{10^{15}h^{-1}M_\odot}}\right)}^{k},
\ee
where the normalization is taken to be $f_{g0}=0.15$, is based on local rich clusters. We look at 
combinations of both mass and redshift dependence for a range of evolutionary
models. In particular, we look at a case of strong evolution given by $k=0.5, s=1$ (Colafrancesco \&
Vittorio, 1994); $k=0.1, s=0.5$ (as suggested by Ettori and Fabian, 1999); $k=0.1, s=0.1$ (weak
evolution); $k=0$ (no mass dependence) and $s=0$ (no redshift dependence).

For the core radius $r_c$ and the temperature, we use 
\bea
r_c\left(\Omega_0, M, z\right) &=& {{1.69h^{-1}Mpc}\o{p}} {{1}\o{1+z}} \times \\ \nonumber
&& \left[ \left( {{M}\o{10^{15}h^{-1}M_\odot}} \right) {{178}\o{\Omega_0 \Delta_c}} \right]^{1/3} ,
\eea

\be
k_B T_e = 7.76 \beta^{-1} {\left({{M}\o{10^{15}h^{-1}M_\odot}}\right)}^{2/3} (1+z) KeV.
\ee 
Here, $\Delta_c(z)$ is the cluster overdensity relative to the background.

Putting everything in, we have the temperature distortion to be 
$\frac{ {\mit\Delta}T(\theta)}{T_{\rm cmb}} = g(x)y(\theta)$, with 

\be
 y(\theta) =
\left(\frac{\sigma_T n_{e0}r_{\rm c}k_{\rm B} T_e}{m_ec^2}\right)   
 \times \frac{\pi e^{1/\xi p^2}}{\sqrt{1+\left(\theta/\theta_{\rm c}\right)^2}}
    {\rm Erfc}\left(
              \sqrt{\frac{1+(\theta/\theta_{\rm c})^2}
                         {\xi p^2}}
              \right).
\ee
The angular core radius $\theta_c = r_c/ d_{ang}$.
The spectral form of the thermal SZ effect is given by
\be
g(x) = {{x^4 e^{x}}\over{{(e^{x} - 1)}^2}}
\left[ x coth(x/2) - 4 \right] ,
\ee
where $x=h\nu/k_BT_{cmb}$. 
This specific spectral dependence of the thermal SZ effect can be used to
separate it out from other CMB anisotropies (Cooray \etal , 2000).

\section{Computing the power spectrum}

The fluctuations of the CMB temperature produced by SZE can be quantified by their spherical harmonic
coefficients $a_{lm}$, which can be defined as
$\Delta T({\bf n}) = T_0^{-1} \sum_{lm} a_{lm} Y_{lm}({\bf n})$. The
angular power spectrum of SZE is then given by
$C_l=<{|a_{lm}|}^2>$, the brackets denoting an ensemble average. 
The power spectrum for the
Poisson distribution of objects, can then be written as 
(Cole \& Kaiser 1988, Peebles 1980)
\be
C_l^{Poisson} = \int_0^{z_{max}} dz {{dV(z)}\over{dz}} \int_{M_{min}}^{M_{max}}
     dM {{dn(M,z_{in})}\over{dM}} {|y_l(M,z)|}^2 ,
\ee	 
where $V(z)$ is the comoving volume and $dn/dM$ 
is the number density of objects.
 
Since these fluctuations occur at
very small angular scales, we can use the small angle approximation of Legendre
transformation and write $y_l$ as the angular Fourier transform of $y(\theta)$
as  $y_l = 2\pi \int y(\theta) J_0 [(l+1/2])\theta]\theta
d\theta $ (Peebles 1980, Molnar \& Birkinshaw 2000). 

In addition to Poisson power spectra, one would expect contribution to a 
`correlation power spectrum' from the clustering of the galaxy clusters.
Following Komatsu and Kitayama (1999), 
we estimate the clustering angular 
power spectrum as
\bea
C_l^{Clustering} &=& \int_0^{z_{max}} dz {{dV(z)}\over{dz}} P_m \times \\ \nonumber
& &{\left[\int_{M_{min}}^{M_{max}} dM {{dn(M,z_{in})}\over{dM}} 
b(M,z_{in}) y_l(M,z)\right]} ^2 ,
\eea
where $b(M,z)$ is the time dependent linear bias factor. The matter power 
spectrum, $P_m(k,z)$, is related to the power spectrum of cluster correlation function 
$P_c(k,M1,M2,z)$ through the bias, i.e
$P_c(k,M1,M2,z)=b(M1,z)b(M2,z)D^2(z)P_m(k,z=0)$ where we 
adopt $b(M,z)$ given by
$b(M,z) = (1+0.5/\nu^4)^{0.06-0.02n}(1 + (\nu^2 -1)/\delta_c)$ (Jing 1999 for details).
This expression
for the bias factor matches accurately the results of
 N-body simulations for a wide range in mass.
In the above equation $D(z)$ is the linear growth factor of density
fluctuation, $\delta_c=1.68$ and $\nu=\delta_c / \sigma(M)$.

\section{Results and Discussions}

We study the power spectrum of SZE from clusters of galaxies, under the assumption of a `precise' and `a priori'
knowledge of
the cosmological parameters. 
We also assume that in the $l$-range of relevance, thermal 
SZE from clusters of galaxies are the
dominant contributors to the temperature anisotropy. The other secondary anisotropies are either 
smaller in strength or contribute at even higher $l's$ or have different spectral dependence (Aghanim
\etal, 2000; Majumdar \etal, 2000). 

We have plotted the Poisson SZE  power spectrum in Fig1, left panel. Clearly,
the primary feature distinguishing a non-evolutionary constant \gmf model from an evolutionary one is the 
position of the peak. The model with a constant \gmf peaks at a higher $l$-value and also has greater power.
The constant \gmf
model peaks at $l\sim4000$. This result is in agreement with that of Komatsu \& Kitayama (1999).
If one assumes that there is no evolution of \fg with redshift (i.e s=0), the peak is at
$l\sim 1100$, whereas in the case of no dependence on mass (k=0), the peak is at $l\sim 2500$. Based 
on EMSS
data (David \etal, 1990), Colafrancesco \& Vittorio (1994) (and also Molnar \& Birkinshaw 2000) model 
\gmf with
k=0.5 \& s=1. For this case, we see that the turnover is at a very low $l\sim 900$. Assuming a mild
evolution (k=0.1, s=0.1), we get the peak at $l\sim 2100$. We also show results for (k=0.5, s=0.5) and (k=0.1,
s=0.5). The last parametrization is based on the recent analysis of $ROSAT$ data by Ettori \& Fabian (1999).
It is evident that the difference, in the $l$-value of the peak of the constant
\gmf scenario from an evolutionary one, can range between $l \sim 1500 - 3200$.. 
The position of the peak thus is a 
strong distcriminatory signature of any evolution of \gmf.

It is easy to understand the shift in the peak of the SZE power spectrum. Let us consider the case s=0, i.e. 
\gmf depends only on total mass. From Eqn.(1), this means an enhanced reduction of \gmf of smaller mass clusters
relative  to the 
larger masses and so a reduction of  power at larger $l's$, (since smaller masses contribute at larger
$l$). Hence, the peak shifts to a lower $l$. For the case k=0, (i.e only redshift dependence),
we now have structures at high $z$ contributing less to the power (than without a redshift dependence). 
Since from PS formalism, less massive structures 
are more abundant at high $z$, this negative dependence of \gmf on redshift cuts off their contribution more 
than the more massive clusters. Hence, once again there is less power at high $l$ and the peak
shifts to lower $l$-value. The parametrization of Eqn. (1) affects the larger masses less, as evident from 
almost equal power seen at $l \lsim 600$, for all models.
The net effect is a reduction of power at smaller angular scales, and 
hence a shift in the position of the peak to a smaller multipole value.

We note that, these results are irrespective of the
arguments given (see Rines \etal, 1999) to explain any possible evolution of \gmf. In their case,
they {\it assume} \gmf to be constant and relies on the cosmology to change the angular diameter distance, so
that there is an `apparent' change in \gmf. In such a case, if there is `actually' even a slight evolution
of \gmf, then one can still account for it with a non-evolutionary model, by simply changing the cosmological
parameters. Our method {\it does not assume a priori} any constancy (or evolution) of \gmf and tries to look for it. 

In Fig1, right panel, we show the SZE clustering power spectrum. For all models, it falls of at a smaller $l$
w.r.t Poisson power spectrum. Since for clustering, the peak depends on the average inter-cluster separation, which 
is fixed once the cosmology is fixed, there is no appreciable spread of the peaks in $l$-space.
The only difference is in their relative
power w.r.t each other which depends on the total gas mass available to distort the CMB.
Addition of the clustering power spectrum to the Poisson case results in slight shift of the peaks to lower
$l$'s.    

It maybe possible to measure the power spectrum of SZE with the ongoing and future high angular resolution
CMB observations. In principle, observations with SUZIE, OVRO, BIMA and ATCA can probe the range in $l$ from $\approx
1000 - 7000$ and a frequency range of $\approx 2 - 350$ GHz. The SZE power spectrum would also be measured with
increased precision by the proposed  ALMA 
and AMiBA (which is geared for blank sky surveys).   

Finally, let us comment on the validity and robustness of our results.
In Fig2, we show results for an open universe ($\Omega_0=0.35, h=0.65$). It is clearly seen that the
difference in the peak position of constant \gmf and evolutionary models remain far apart (Infact, for 
same parameters of $k=0.5, s=1$, the difference increases to $\approx 4500$ from that of
$\approx 3000$ in a flat universe). 
It is seen that the turnover of the SZE power spectrum is insensitive to the mass cutoff, since  
main contribution to the  anisotropy comes from clusters with 
$10^{14}M_\odot <M<10^{15}M_\odot$.
In Fig2, we also indicate the effect of having a more compact gas distribution with $p=7$. 
We see that shift in the peaks are negligible (though the height is reduced a little) and an
uncertainty as to how far the gas extends is not major. 
The use of a single $\beta$ to model the full gas distribution introduces little error, though 
a $\beta$-model fits the inner cluster regions better.
 This is because the major contribution to the anisotropy comes from around the 
core region, and increasing $\beta$ slightly decreases the overall distortion, without touching the peak.
Also, a modified  $M-T$ relation (more suitable for $\Lambda$CDM) does not change the conclusions of this paper
(although amplitude of distortion slightly changes). 
For a more detailed analysis, however, one should take better observationally supported gas density and temperature profiles
(see Yoshikawa \& Suto, 1999). These points will be discussed in greater detail in a future publication. 

In conclusion, we have computed the angular power spectrum of SZE from clusters of galaxies.
We have shown the position of the peak of the power 
spectrum to bear a {\it strong discriminatory signature} of different \gmf models. One of the goals of arc minute scale 
observations of the CMB
anisotropy is to measure the SZE power spectrum from blank sky surveys. Such observational results can be used to constrain
\gmf models. This also  has the added advantage of being devoid of uncertainties that can creep in through `selection biases' in
estimating the \gmf using pointed studies of X-ray selected galaxy cluster. Our method, thus, provides a powerful
probe of evolution (or constancy) of gas mass fraction and can potentially resolve the decade long debate.

\acknowledgments
The author wishes to thank Joe Mohr and Biman Nath for critical comments, Sunita Nair and Dipankar Bhattacharya for
valuable suggestions and Pijush Bhattacharjee for encouragements. He also wishes to thank the Raman Reseach Institute,
Bangalore. After the completion of this work, we learnt about a complimentary work 
by Carlstrom \& Holder (2001) which looks at changes in gas profiles due to preheating and its effect on SZE.


\beb

\bi Aghanim, N., Balland, C., \& Silk, J., 2000, A\&A, 357, 1
\bi Allen, S. W., \& Fabian, A., 1998, MNRAS, 297, L57
\bi Bardeen, J. M., Bond, J. R.,Kaisaer, N. \& Szalay, A. S., 1986
ApJ, 304. 15
\bi Bialek, J. J., Evrard., A. E., \& Mohr, J, 2000, preprint, (astro-ph/0010584)
\bi Birkinshaw, M. 1999;Phys. Rep., 310, 97
\bi Bunn, E. \& White, M. 1997, ApJ, 480, 6
\bi Carlstrom, J. E., \& Holder, G., 2001, private communication
\bi Colafrancesco, S., \& Vittorio, N., 1994, ApJ, 422, 443 
\bi Cole, S. \& Kaiser, N., 1988, MNRAS, 233, 637
\bi Cooray, A., Hu, W., \& Tegmark, M., 2000, preprint, (astro-ph/0002238)
\bi David, L. P., Arnaud, K. A., Forman, W., \&  Jones, C., 1990. ApJ, 356, 32
\bi David, L. P., Jones, C., \& Forman, W., 1995, ApJ, 445, 578
\bi Ettori, S., \& Fabian, A. C., 1999, MNRAS, 30. 834
\bi Grego, L., \etal, 2000, preprint, (astro-ph/0012067)
\bi Holder, G. P., \& Carlstrom, J. E., 1999, preprint, (astro-ph/9904220)
\bi Hu, W., Fukugita, M., Zaldarriaga, M., \& Tegmark, M., 2000, preprint, (astro-ph/006436)
\bi Jing, Y. P., 1999, ApJ, 515, L45
\bi Komatsu, E. \& Kitayama, T. 1999, ApJ, 526, L1 
\bi Majumdar, S., Nath, B. B., \& Chiba, M., 2000, preprint, (astro-ph/0012016)
\bi Mohr, J., Mathiesen, B., \& Evrard, A., 1999, ApJ, 517, 627
\bi Molnar, S. M. \& Birkinshaw, M., 2000, ApJ, 537, 542
\bi Padmanabhan, T., \& Sethi, S., 2000, preprint, (astro-ph/0010309)
\bi Peebles, P. J. E. 1980;The Large Scale Structure of the
Universe;Princeton;Princeton Univ. Press;
\bi Ponman, T. J., Bourner, P. D. J., Ebeling, H., \& Bohringer, H., 1996, MNRAS, 283, 690
\bi Press, W. H., \& Schechter, P. 1974, ApJ, 187, 425
\bi Refregier, A., Komatsu, E., Spergel, D. N., \& Pen, U, 1999, preprint, (astro-ph/9912180)
\bi Rines, K., Forman, W., Pen, U., Jones, C., \& Burg, R., 1999, ApJ, 517, 70
\bi Sadat, R. \& Blanchard, A., 2001, preprint (astro-ph/0102010)
\bi Sasaki, S., 1996, PASJ, 48, L119
\bi Schindler, S., 1999, A\&S, 349, 435
\bi Sheth, R. K., \& Tormen, G., 1999, MNRAS, 308, 119
\bi Sugiyama, N., 1995, ApJS, 100, 281
\bi Tsuru T.G., \etal, 1997, preprint, (astro-ph/9711353)
\bi White, D. A., \& Fabian, A. C., 1995, MNRAS, 273, 72
\bi Yoshikawa, K.,  \& Suto, Y., 1999, ApJ ,513, 549
\bi Zel'dovich, Ya. B., \& Sunyaev, R. A. 1969, Ap\&SS 4, 301;

\eeb


\clearpage
\begin{figure}
\plottwo{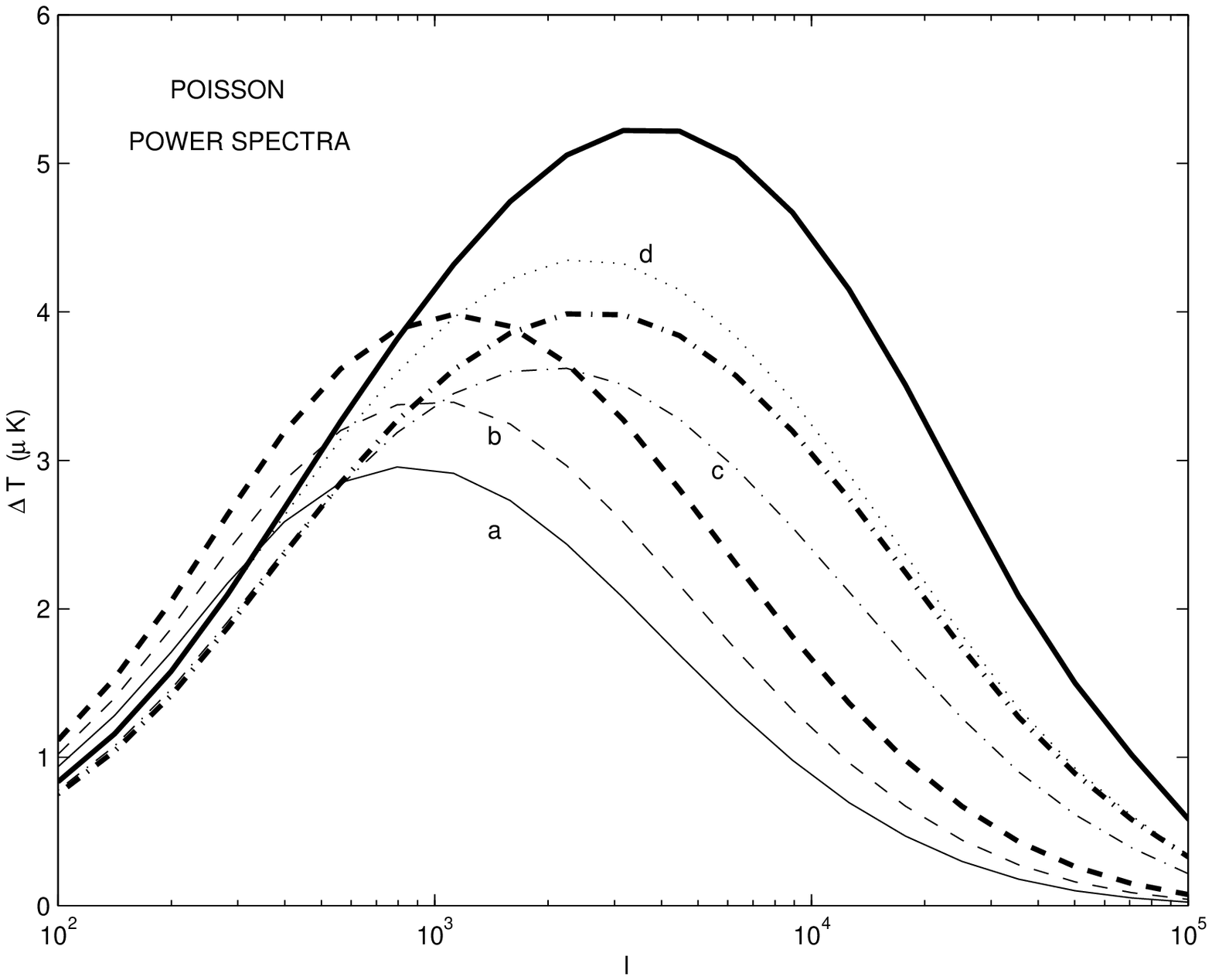}{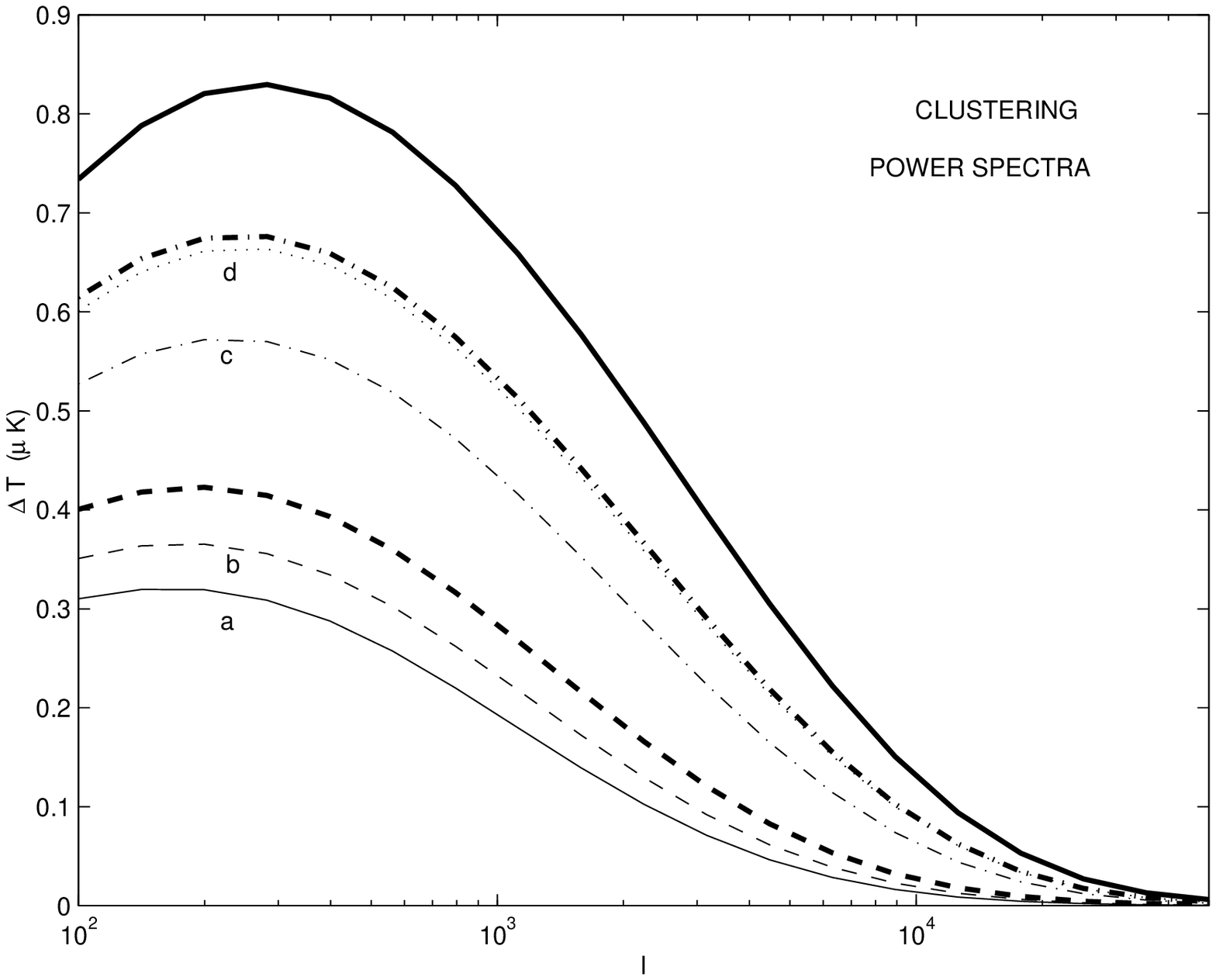}
\caption{
The Poisson (left panel) and clustering (right panel) power spectra due to SZE from galaxy 
clusters for different \gmf models. For both the panels, the thick solid line correponds to constant \gmf model, the
thick dashed line has no evolution with redshift and the thick dash-dotted line has no evolution with total mass.
The thin lines are for the cases: a) k=0.5, s=1; b) k=0.5, s=0.5; c) k=0.1, s=0.5 and d) k=0.1, s=0.1.
\label{fig1}
}
\end{figure}

\clearpage
\begin{figure}
\plotone{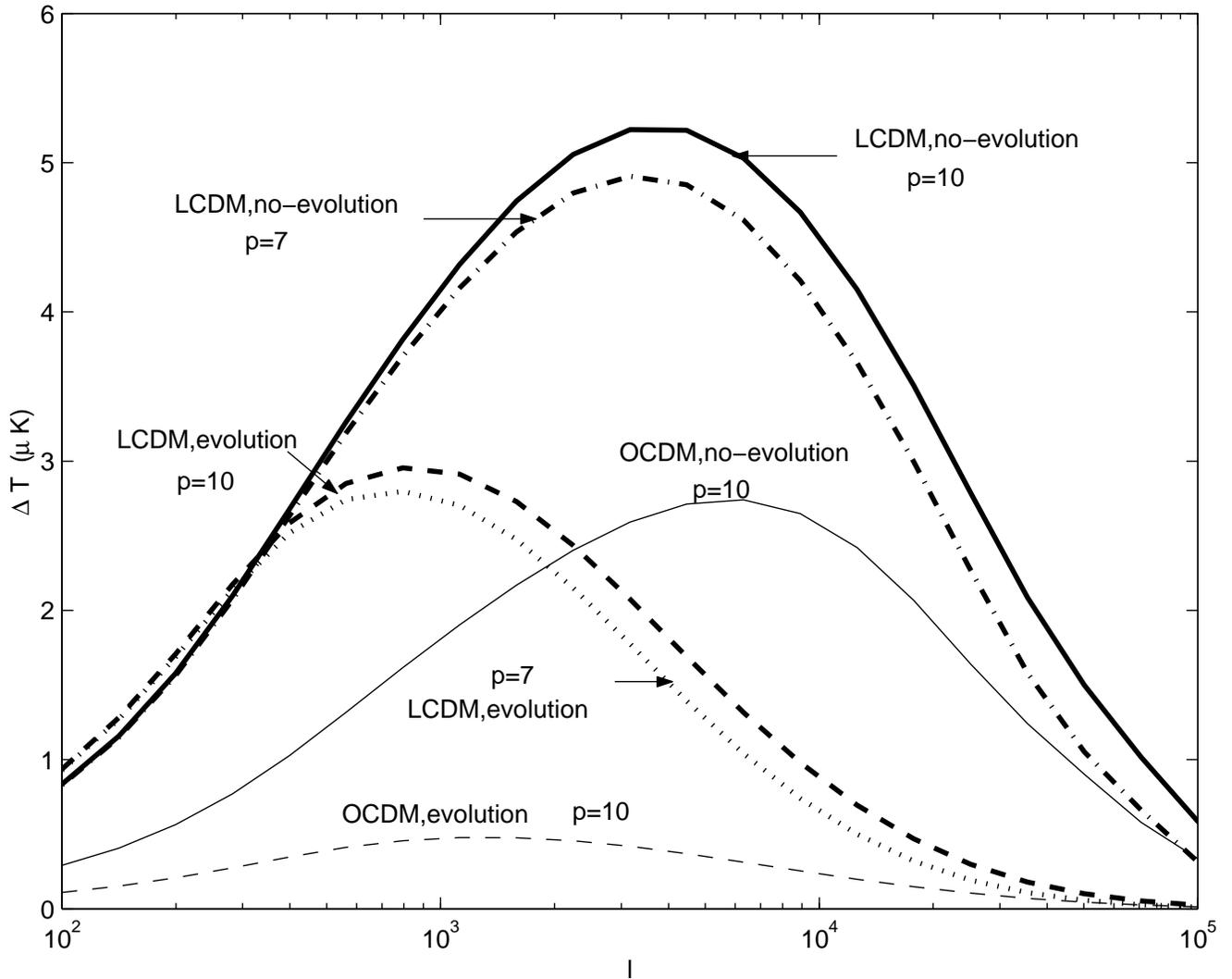}
\caption{
The Poisson SZE power spectra are plotted for different cosmologies and with different extension of the 
gas mass. The solid lines are for a $\Lambda CDM$, with $\Omega_m=0.35, \Omega_\Lambda=0.65, h=0.65$, and the thin
lines are for $OCDM$ with $\Omega_m=0.35,h=0.65$. The $OCDM$ lines have been multiplied by a factor of 10 in the
plot. The solid and dashed lines are for gas mass extending upto 
$10 r_c$, whereas the dash-dotted and the dotted lines are for extension upto $7 r_c$.
\label{fig2}
}
\end{figure}

\end{document}